\documentclass[aps,floats]{revtex4}
\usepackage{tabularx,graphicx}
\usepackage{epsfig}
\usepackage{amsmath}
\usepackage{bbm}
\usepackage{graphicx}

\begin{document}

\title{Rippling transition from electron-induced condensation of curvature field in graphene}
\author{J. Gonz\'alez  \\}
\address{Instituto de Estructura de la Materia,
        Consejo Superior de Investigaciones Cient\'{\i}ficas, Serrano 123,
        28006 Madrid, Spain}

\date{\today}

\begin{abstract}

A quantum field theory approach is applied to investigate the dynamics of flexural phonons 
in a metallic membrane like graphene, looking for the effects deriving from the strong interaction 
between the electronic excitations and elastic deformations. Relying on a self-consistent
screening approximation to the phonon self-energy, we show that the theory has a critical 
point characterized by the vanishing of the effective bending rigidity of the membrane at low
momentum. We also check that the instability in the sector of flexural phonons takes place without
the development of an in-plane static distortion, which is avoided due to the significant reduction 
of the electron-phonon couplings for in-plane phonons at large momenta. Furthermore, we analyze
the scaling properties of the many-body theory to identify the order parameter that opens up at the
point of the transition. We find that the vanishing of the effective bending rigidity and the onset 
of a nonzero expectation value of the mean curvature are 
concurrent manifestations of the critical behavior. The results presented here imply that, even in 
the absence of tension, the theory has a critical point at which the flat geometry becomes an unstable 
configuration of the metallic membrane, with a condensation of the mean curvature field that may well 
reproduce the smooth distribution of ripples observed in free-standing graphene.

\end{abstract}

\maketitle



\section{Introduction}

During the last decade many efforts have been devoted to the investigation of graphene\cite{geim}, the material consisting of a one-atom-thick layer of carbon atoms with outstanding electronic, elastic and optical properties\cite{rmp}. These derive to a great extent from the reduced dimensionality of the material. From the point of view of the elastic behavior, the carbon sheet has shown to be a very unusual kind of membrane, placed in a regime where anharmonic effects are very important. Regarding the electronic properties, graphene displays also unconventional features arising from the relativistic-like dispersion of the valence and conduction bands, that touch only at isolated points in the corners of the Brillouin zone. 

The interplay between the electronic excitations and elastic degrees of freedom in graphene has been investigated in recent years, counting on the fact that the electron-phonon couplings should be rather large in the carbon layer\cite{efe}. Most part of the studies have concentrated at first on the effect of the elastic modes on the electronic behavior. Focusing on the out-of-plane deformations (so-called flexural phonon modes), it has been found that they can have a significant impact on the electronic transport properties of the material\cite{kg,mar,castro}. Considering the interaction with the substrate, it has been also shown that the corrugation induced on the carbon sheet may lead to local changes in the Fermi energy, making graphene to behave as a metallic membrane susceptible to develop charge puddles\cite{kim}. There have been also other studies stressing the potential of the irregularities in the shape of the membrane to modify the band structure of the material\cite{hor}.

Conversely, there have been recent works devoted to investigate the effect of the electronic degrees of freedom on the dynamics of the flexural phonons\cite{gazit,prl,gdw}. The main aim of these investigations is the explanation of the formation of ripples in graphene. In this respect, the carbon layer is known to have corrugation, which is mainly induced in some cases by the substrate\cite{stol,ger}, but ripples have been also observed in suspended samples of the material\cite{meyer}. In Ref. \cite{prl}, it has been shown that the coupling to the electron-hole polarization may give rise to a significant renormalization of the elastic constants of the metallic membrane. For sufficiently strong deformation potential (electron-phonon coupling), it has been found that the bending rigidity undergoes a very drastic reduction, leading to a critical point and spontaneous symmetry breaking in the system of flexural phonons. A different approach has been proposed recently in Ref. \cite{gdw}, where the hybridization of electron-hole and flexural phonon excitations is investigated in the strong coupling regime, finding the simultaneous development of charge puddles and spontaneous breakdown of symmetry of the out-of-plane deformations. 

In these studies of the dynamics of flexural phonons, the critical point is found for values of the deformation potential in the range of $25-30$ eV. This point marks the instability of the flat geometry of the membrane, but the relatively large values of the electron-phonon coupling oblige to check that a Kohn anomaly with the consequent in-plane static distortion does not arise before the symmetry breaking in the sector of flexural phonons. This analysis requires in turn some knowledge about the behavior of the electron-phonon couplings when going from the long-wavelength limit to the regime of large momenta. 

Moreover, the symmetry-breaking pattern giving rise to ripple formation may depend on the balance between different order parameters. In Ref. \cite{prl}, the condensation was found to be related to the gaussian curvature of the membrane, with a mechanism very similar to that opening a nonvanishing expectation value for a Higgs field. In that approach, the spontaneous symmetry breaking is driven by any slight amount of negative tension, being however much unclear the state of the membrane in the limit situation where that parameter vanishes. This raises the question about the dominant order parameter of the instability when the metallic membrane is not under the influence of any external perturbation.

In this paper we address these last two questions, applying self-consistent methods in the renormalization of the many-body theory of flexural phonons. The general framework goes back to the pioneering work of Ref. \cite{np}, which considered the problem of the stability of a crystalline membrane in the classical regime. For this kind of statistical system, the renormalization of elastic constants was further refined in Ref. \cite{dr} with the use of a self-consistent screening approximation (see also \cite{gazit2}). This computational scheme has proven to be very appropriate for the study of the many-body properties of flexural phonons\cite{rold1}, whose interactions account for important anharmonic effects in graphene\cite{k1,k2,rold2}.

Going one step beyond, we will consider the quantum statistical formulation of the many-body theory of flexural phonons, which is pertinent for a stiff membrane like graphene where quantum fluctuations may be relevant even for not very low temperatures. In this framework, we will compute the flexural phonon self-energy by means of a self-consistent screening approximation, in a theory where the effective interaction between flexural phonons encodes the coupling to the electron-hole excitations as well as to in-plane phonons. We will incorporate in particular the momentum dependence of the electron-phonon couplings for a sensible determination of the critical point of the metallic membrane, characterized by the vanishing of the effective bending rigidity at a low value of the momentum. 

We will also identify the order parameter that arises naturally at the critical point. When the effective bending rigidity is drastically reduced, the flexural modes become very soft at low momenta, making the system susceptible to undergo condensation with a definite content of phonon fields. We will see that the mean curvature (the laplacian of the flexural phonon field) is the field that gets a nonvanishing expectation value at the point of the transition, playing therefore the role of order parameter. This complements the analysis of Ref. \cite{prl} showing that, even in the absence of tension, there is a natural condensation mechanism leading to the destabilization of the flat geometry of the metallic membrane. In this respect, the mean curvature is a quantity clearly constrained on the average in a planar geometry, which does not prevent however that the transition may take place with the formation of spatial domains with alternating sign of the order parameter, producing a landscape that may well reproduce the smooth distribution of ripples observed experimentally in graphene samples.

\section{Effective action of flexural phonons}

In the case of graphene considered as a metallic membrane, there is a clear hierarchy regarding the different degrees of freedom of the system. The energy scale of the electronic excitations is much higher than that of the in-plane phonons, that are found in turn at energies above those of the flexural phonons. Thus, in order to obtain an effective theory for the latter, it is pertinent to integrate out first the fields of the electron quasiparticles, carrying out the same task subsequently for the in-plane phonons. 

We begin then with the system of electron quasiparticles, that we represent in terms of a set of two-component Dirac fields $\psi_n ({\bf r})$ accounting for valley and spin degrees of freedom, coupled to the elastic deformations of the membrane. These are described by the displacement within the reference plane ${\bf u} ({\bf r}) = (u_1 ({\bf r}), u_2 ({\bf r}))$ and the vertical shift $h ({\bf r})$, which build together the strain tensor
\begin{equation}
u_{ij} = \frac{1}{2}  (\partial_i u_j + \partial_j u_i + \partial_i h  \partial_j h)
\label{st}
\end{equation} 
Assuming that the electronic correlations are weak in graphene, we may characterize the dynamics of the electron quasiparticles by means of a quadratic Dirac hamiltonian. The action including the electron-phonon coupling can be written in the form
\begin{equation}
S_e = \int dt \:d^2 r \; \psi^{\dagger}_n({\bf r}) 
  \left (i\partial_t - iv_F\mbox{\boldmath $\sigma $}  \cdot  \mbox{\boldmath $\nabla $} \right)
     \psi_n ({\bf r})   
     - g \int dt \: d^2 r  \: \psi^{\dagger}_n({\bf r})  \psi_n ({\bf r}) \: {\rm Tr}  \left( u_{ij} ({\bf r}) \right)
\label{acte}
\end{equation}
where $v_F$ is the Fermi velocity and $g$ stands for the so-called deformation potential (measured in energy units). 

Upon integration of the Dirac fields $\psi_n ({\bf r})$, we get a contribution to the effective action that depends on $u_i ({\bf r})$ and $h ({\bf r})$. While this generates a sequence of terms with increasing powers of those fields, particular attention has to be paid to the contribution proportional to the square of the strain tensor, since it gives rise to a direct renormalization of the couplings for the elastic deformations. The original action for the elastic degrees of freedom is
\begin{equation}
S_{\rm el} = \frac{1}{2} \int dt \: d^2 r \left( \rho  (\partial_t \mathbf{u} )^2 + \rho  (\partial_t h )^2 - \kappa_0 (\nabla^2 h)^2 - 2\mu {\rm Tr}(u_{ij}^2) - \lambda  ({\rm Tr}(u_{ij}))^2  \right)
\label{s1}
\end{equation}
where $\rho $ is the mass density, $\kappa_0 $ is the bending rigity, and $\mu, \lambda$ stand respectively for the in-plane shear and bulk moduli. Upon integration of the fermion fields, we obtain a correction to $S_{\rm el}$
\begin{equation}
\Delta S_{\rm el} = - \frac{1}{2} g^2 \int dt \: d^2 r   \int dt' \: d^2 r' \left. ( \partial_i u_i + \frac{1}{2} \partial_i h \partial_i h )\right|_{{\bf r}, t} \chi ({\bf r}-{\bf r}', t-t') \left. ( \partial_j u_j + \frac{1}{2} \partial_j h \partial_j h )\right|_{{\bf r}', t'}
\end{equation}
where the new interaction is mediated by the electron-hole polarization $\chi ({\bf r}, t)$. 

For graphene with $N = 4$ Dirac fermion flavors, we have the Fourier transform $\chi ({\bf q}, \omega_q ) = -{\bf q}^2/4\sqrt{v_F^2 {\bf q}^2 - \omega_q^2}$. As long as the typical energies of phonons (in-plane as well as out-of-plane) are much lower that those of the electronic excitations, it is a good approximation to set $\omega_q = 0$ in $\chi ({\bf q}, \omega_q )$ for the evaluation of the effective action $S_{\rm eff} = S_{\rm el} + \Delta S_{\rm el}$. This can be written then in frequency and momentum space as
\begin{eqnarray}
S_{\rm eff} & = &  \frac{1}{2} \int \frac{d^2 q}{(2\pi)^2} \frac{d\omega}{2\pi} \; (\rho \: \omega^2  -
\kappa_0 {\bf q}^4  ) h({\bf q}, \omega) \: h(-{\bf q}, -\omega)
                                                      \nonumber                          \\
  &  &  +  \frac{1}{2} \int \frac{d^2 q}{(2\pi)^2} \frac{d\omega}{2\pi}  \;
 (  \rho \: \omega^2 \: u_i({\bf q}, \omega)  u_i (-{\bf q}, -\omega) 
  - 2\mu \: u_{ij}({\bf q}, \omega)  u_{ij} (-{\bf q}, -\omega)
  - \lambda'({\bf q}) \: u_{ii}({\bf q}, \omega)  u_{jj} (-{\bf q}, -\omega)  )
\end{eqnarray}
with $\lambda'({\bf q}) = \lambda - g^2 |{\bf q}| /4v_F$, and $u_{ij}({\bf q}, \omega)$ being the Fourier transform of the strain tensor (\ref{st}).
 
We can integrate out the in-plane phonons at this point, using a known trick that consists in separating the transverse component of the strain tensor\cite{np}. Introducing the transverse projector $P_{ij}^{(T)} = \delta_{ij} - (\nabla^2)^{-1} \partial_i \partial_j$, we can recast the strain tensor in the form
\begin{equation}
u_{ij} = \frac{1}{2} \left (\partial_i \bar{u}_j + \partial_j \bar{u}_i + P_{ij}^{(T)} (P_{kl}^{(T)}\partial_k h  \partial_l h) \right)
\end{equation} 
Given that the energy scale of the in-plane phonons is higher than that of the flexural phonons, we can take again a static approximation to study the dynamics of the latter in the background of in-plane phonons. Then, only the longitudinal part of the field $\bar{u}_i$ couples to $P_{kl}^{(T)}\partial_k h  \partial_l h$, and we get an expression for the effective action of the $h$ field
\begin{eqnarray}
S_{\rm eff}  & = &   \frac{1}{2} \int \frac{d^2 q}{(2\pi)^2} \frac{d\omega}{2\pi} \; (\rho \: \omega^2  -
\kappa_0 {\bf q}^4  ) h({\bf q}, \omega) \: h(-{\bf q}, -\omega)
                                                 \nonumber                  \\
  &  &  -  \frac{1}{8} \int \frac{d^2 q}{(2\pi)^2} \frac{d\omega}{2\pi}  \;
K({\bf q}) \: P_{ij}^{(T)} ({\bf q}) \: h_{ij} ({\bf q}, \omega) \: P_{kl}^{(T)} ({\bf q}) \: h_{kl} ({-\bf q}, -\omega)
\label{act}
\end{eqnarray}
with 
\begin{equation}
K({\bf q}) = 2\mu + \lambda - g^2 |{\bf q}|/4v_F
 - \frac{(\lambda - g^2  |{\bf q}|/4v_F)^2 }
                               {2\mu + \lambda - g^2 |{\bf q}|/4v_F }
\label{kq}
\end{equation}
and $h_{ij}({\bf q}, \omega)$ standing for the Fourier transform of $\partial_i h \partial_j h$.

The action (\ref{act}) constitutes our starting point to study the effects of the interaction of the flexural phonons in the metallic membrane. It can be actually seen that it defines a scale invariant field theory, since its expression remains unmodified (except for the scaling of the electron-phonon coupling) under the change of variables
\begin{equation}
\omega' = s \omega  \;\; , \;\;  {\bf q}' = \sqrt{s} {\bf q}  \;\; , \;\; 
           h'({\bf q}', \omega') = \frac{1}{s^2} h ({\bf q}, \omega)
\label{scale}
\end{equation}
It makes sense therefore to investigate the scaling of the theory at the quantum level. This leads to characterize the effective energy dependence of the parameters in the action, whose low-energy behavior allows in turn to get information about the main elastic properties of the membrane in the long-wavelength limit.

\section{Self-consistent screening approximation}

For sufficiently large values of the deformation potential $g$, the coupling function
$K({\bf q})$ may reach negative values, driving the interaction into a regime
of attraction for a certain range of the momentum ${\bf q}$. As we are going to see, this
may lead to an instability in the metallic membrane, either from the decay of the flexural 
phonons or from a drastic renormalization of the bending
rigidity. In order to claim the instability in the sector of flexural phonons, we have to 
make sure however that the pole in the denominator of $K({\bf q})$ is not hit for any allowed 
momentum, since that would imply a different instability pertaining to the in-plane phonons. 

The last term at the right-hand-side of (\ref{kq}) represents actually the contribution to 
the coupling function from the exchange of in-plane phonons, and reaching the pole in their 
propagator would give rise to a large Kohn anomaly implying the formation of 
an in-plane static distortion. Before that happens as the value of $g$ is increased, it 
is possible however to find the instability in the sector of flexural phonons at a lower 
value of the deformation potential, signaling the spontaneous breakdown of symmetry in the 
metallic membrane but with regular behavior of the in-plane phonons. 

\begin{figure}[h]
\begin{center}
\mbox{\epsfxsize 2.0cm \epsfbox{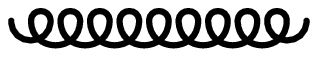}} \hspace{0.5cm} {\Large $=$}
 \hspace{0.5cm}  \mbox{\epsfxsize 2.0cm \epsfbox{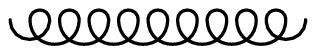}}
 \hspace{0.5cm}  {\Large $+$}  \hspace{0.5cm}
\mbox{\epsfxsize 4.8cm \epsfbox{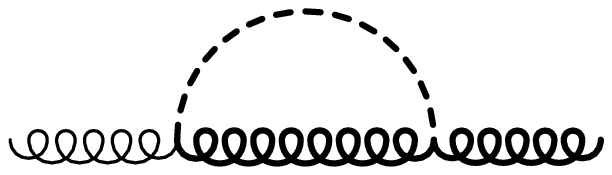}}  \\  \vspace{1cm}
\raisebox{0.2cm}{\epsfxsize 2.0cm \epsfbox{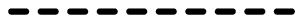}} \hspace{0.5cm} 
  \raisebox{0.16cm}{\Large $=$}
 \hspace{0.5cm}  \raisebox{0.2cm}{\epsfxsize 2.0cm \epsfbox{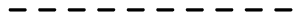}}
 \hspace{0.5cm}  
  \raisebox{0.16cm}{\Large $+$} \hspace{0.5cm}  
    \raisebox{-0.8cm}{\epsfxsize 4.8cm \epsfbox{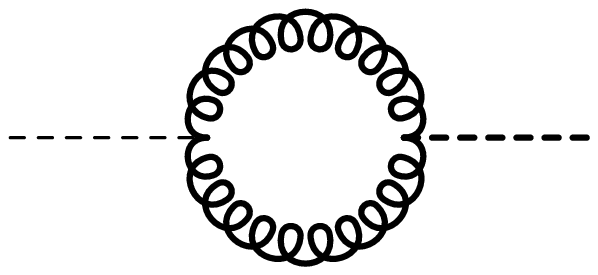}} 
\end{center}
\caption{Diagrammatic equations defining the self-consistent screening approximation. The thick(thin) curly line represents the dressed(free) flexural phonon propagator and the thick(thin) dashed line represents the dressed(undressed) flexural phonon interaction.}
\label{zero}
\end{figure}

We study these effects using a self-consistent screening approximation, which can be 
defined by the diagrammatic equations shown in Fig. \ref{zero}. The propagator 
$D({\bf q}, \omega_q)$ of the flexural phonon interaction is dressed in the form
\begin{equation}
D^{-1}({\bf q}, \omega_q) =  K^{-1}({\bf q}) +  \Pi ({\bf q}, \omega_q)
\end{equation}
with the polarization given in terms of the flexural phonon propagator 
$G ({\bf k}, \omega_k)$ as
\begin{equation}
i\Pi ({\bf q}, \omega_q) = \frac{1}{2} 
        \int \frac{d^2 k}{(2\pi)^2}  \frac{d\omega_k }{2\pi}
\left( {\bf k}^2 - \frac{({\bf q} \cdot {\bf k})^2}{{\bf q}^2} \right)^2   \:
G ({\bf k}, \omega_k)  \:   G ({\bf q}-{\bf k}, \omega_q - \omega_k)
\label{pi}
\end{equation}
In turn, we use the dressed interaction to obtain the flexural phonon self-energy 
\begin{equation}
i\Sigma ({\bf q}, \omega_q) = 
  \int \frac{d^2 k}{(2\pi)^2}  \frac{d\omega_k }{2\pi}
  \frac{({\bf q}^2{\bf k}^2 - ({\bf q} \cdot {\bf k})^2)^2}{|{\bf q}-{\bf k}|^4}
  \: D({\bf q}-{\bf k}, \omega_q - \omega_k) \:  G ({\bf k}, \omega_k)
\label{selfen}
\end{equation}
which must give back the phonon propagator according to the equation 
\begin{equation}
G^{-1}({\bf q}, \omega_q) = \rho \omega_q^2 - \kappa_0 {\bf q}^4  + 
            \Sigma ({\bf q}, \omega_q)
\label{self}
\end{equation}

The self-energy $\Sigma ({\bf q}, \omega_q)$ in (\ref{selfen}) becomes 
proportional to ${\bf q}^4$, which means that its main effect is the renormalization 
of the bending rigidity $\kappa_0$. Therefore, we may carry out the self-consistent 
resolution modifying the flexural phonon propagator to account for a dressed 
bending rigidity 
\begin{equation}
G({\bf q}, \omega_q) = 
  \frac{1}{\rho \omega_q^2 - \kappa ({\bf q},\omega_q) \: {\bf q}^4 + i\eta}
\end{equation}

Regarding the interaction, it is crucial to keep the deformation potential below the upper 
limit at which the pole in the last term of $K({\bf q})$ in (\ref{kq}) is reached (which happens 
first for large momenta, of the order of the inverse of the lattice spacing). For a more 
precise evaluation of that regime, it is pertinent to discern the own momentum dependence of the 
electron-phonon couplings which, in the case of the in-plane phonons, turn out to be at large 
momenta much smaller than expected from the nominal value of the deformation potential approached 
at long wavelengths. 

In this analysis, one is led to distinguish between the electron-phonon coupling for in-plane 
phonons appearing in the denominator at the right-hand-side of (\ref{kq}) and the electron-phonon 
coupling for flexural phonons making the direct quadratic contribution to $K({\bf q})$. 
For the self-consistent resolution, we have refined this coupling function using the tight-binding 
expressions for both types of electron-phonon couplings discussed in the Appendix. This approach is 
simple enough to produce manageable expressions of the couplings, while capturing the nonlinear 
behavior at large momenta. Taking for instance for the elastic constants the values 
$\mu\approx 5.7 \mbox{ eV/\AA}^2$ and $\lambda\approx 1.6 \mbox{ eV/\AA}^2$ \cite{sst}, we can 
estimate in this way the limit set by the in-plane instability at a deformation 
potential $g \approx 29$ eV.

In practice, we have resorted to a numerical computation to obtain self-consistently 
$\Pi({\bf q}, \omega_q)$ and $\kappa ({\bf q}, \omega_q)$. We have evaluated the expressions
(\ref{pi}) and (\ref{selfen}) 
by performing a Wick rotation $\omega = i \overline{\omega }$ and summing over a set of discrete 
Matsubara frequencies $\overline{\omega}_n = 2\pi n T$ with $n = 0, \pm 1, \pm 2, \ldots$, which 
corresponds to placing the theory at a finite temperature $T$. In this framework, we have for the 
polarization
\begin{equation}
\Pi ({\bf q}, i\overline{\omega}_q) = \frac{1}{2} \int \frac{d^2 k}{(2\pi)^2}  
   \left( {\bf k}^2 - \frac{({\bf q} \cdot {\bf k})^2}{{\bf q}^2} \right)^2   \:
    T \sum_{\overline{\omega}_n} 
\frac{1}{\rho \overline{\omega}_n^2 + \kappa ({\bf k}, i\overline{\omega}_n) \: {\bf k}^4} \:
\frac{1}{\rho (\overline{\omega}_q-\overline{\omega}_n)^2 + 
    \kappa ({\bf q}-{\bf k},i\overline{\omega}_q-i\overline{\omega}_n) \: ({\bf q}-{\bf k})^4}
\label{pi2}
\end{equation}
On the other hand, the equation for the phonon self-energy becomes a condition for the 
self-consistent renormalization of the bending rigidity 
\begin{equation}
\kappa({\bf q}, i\overline{\omega}_q) = \kappa_0  +  
\int \frac{d^2 k}{(2\pi)^2}  
 \frac{({\bf k}^2 - (\hat{\bf q} \cdot {\bf k})^2)^2}{|{\bf q}-{\bf k}|^4} \:
 T \sum_{\overline{\omega}_n} 
 \frac{D({\bf q}-{\bf k}, i\overline{\omega}_q-i\overline{\omega}_n)}
    {\rho \overline{\omega}_n^2 + \kappa ({\bf k},i\overline{\omega}_n) \: {\bf k}^4}
\label{ka}
\end{equation}
with $\hat{\bf q} = {\bf q}/|{\bf q}|$.

\begin{figure}[h]
\begin{center}
\mbox{\epsfxsize 4.8cm \epsfbox{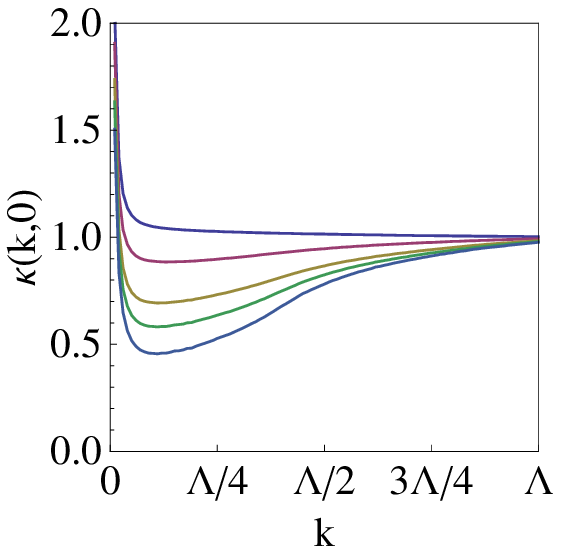} \hspace{3cm}
  \epsfxsize 4.8cm \epsfbox{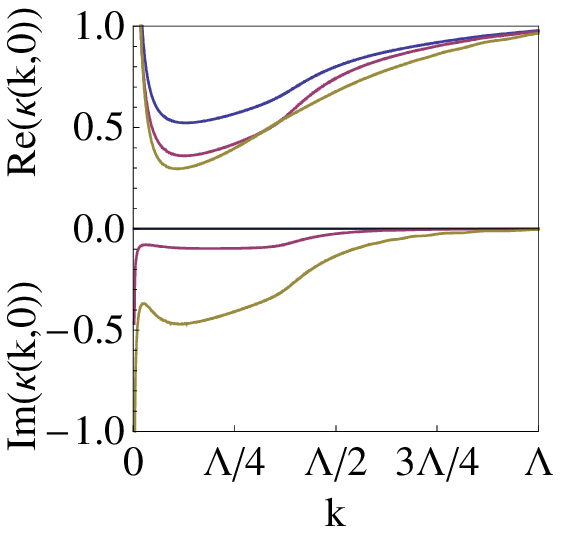}}  \\
 \hspace{0.3cm}  (a) \hspace{6.6cm} (b) 
\end{center}
\caption{(a) Plot of the effective bending rigidity for $2\pi T = 5$ meV and values of the deformation potential $g = 0, 28.0, 28.6, 28.68, 28.72$ eV (curves from top to bottom). (b) Plot of the real and imaginary parts of the effective bending rigidity for $2\pi T = 10$ meV and values of the deformation potential $g = 28.7, 28.74, 28.8$ eV (curves from top to bottom in both sides of the figure). $\Lambda $ stands for the cutoff in the momentum integrals, which is of the order of the inverse of the lattice spacing.}
\label{one}
\end{figure}

The remaining integrals in Eqs. (\ref{pi2}) and (\ref{ka}) can be  
approximated by summing over the points of a grid covering the momentum space,
up to the boundary of the Brillouin zone. Usually, taking 200 divisions in both 
the angular variable and the modulus of the momentum 
leads to results with reasonable precision, assuring also the convergence of 
the recursive resolution of the integral equations within $20-30$ iterations. 
The behavior of the bending rigidity $\kappa({\bf q}, 0)$ obtained with such a 
procedure is represented in Fig. \ref{one}(a), for a temperature such that 
$2\pi T = 5$ meV. It becomes clear the downward renormalization of the bending 
rigidity for increasing values of the deformation potential, leading to significant
deviations from the resolution without electron-phonon interaction for $g \gtrsim 25$ eV.

Following the decreasing trend of $\kappa $, there is however a value of the 
deformation potential at which the iterative resolution of Eqs. (\ref{pi}) and 
(\ref{ka}) becomes unstable for a purely real function 
$\kappa({\bf q}, i\overline{\omega}_q)$. The convergence of the procedure can 
be restored by assuming that the effective bending rigidity has an imaginary part, 
which starts having a nonvanishing value for a deformation potential
$g^* \approx 28.7$ eV (for $2\pi T = 10$ meV). This is shown in Fig. 
\ref{one}(b), which represents the real and imaginary parts of 
$\kappa({\bf q}, 0)$ for different values of the deformation potential
at $2\pi T = 10$ meV.

These results reflect that, before a regime with unphysical negative values of 
$\kappa $ is reached, the flexural phonons have a more conventional instability 
characterized by a nonzero imaginary part of the self-energy. This is the signature 
of a finite lifetime of the phonon modes, which can decay into collective 
excitations of the strongly interacting system. Indeed, it can be shown already
at a perturbative level that single-phonon states are unstable towards the emission
of flexural phonon pairs or more exotic hybrid excitations. We stress anyhow that 
the decay process described here has nonperturbative character, since it arises even 
after performing the Wick rotation devised to circumvent the
pole in the phonon propagator. This explains that the phonon decay rate may have an
unusual behavior in the limit $\omega_q \rightarrow 0, {\bf q} \rightarrow 0$, as 
observed from the plot of Fig. \ref{one}(b). It can be shown that the ratio
${\rm Im} (\kappa({\bf q}, 0))/{\rm Re} (\kappa({\bf q}, 0))$ converges however to a 
finite value in that limit, implying that the decay rate of the low-energy phonon 
modes is proportional to their energy. 

The growth of the imaginary part of $\kappa({\bf q}, 0)$ at low momenta is actually
a consequence of the crossover to the regime dominated by thermal effects. At 
zero temperature, the effective momentum dependence of the bending rigidity 
can be obtained as a renormalization effect characteristic of the quantum theory, which
leads to the low-energy behavior (neglecting the electron-phonon interaction)\cite{fn}
\begin{equation}
\kappa \sim  |\log ({\bf q})|^{4/7}
\end{equation}
As $T \neq 0$, however, there is always some regime, corresponding to phonon energies much 
smaller than the thermal energy, where the thermal effects take over, changing the behavior 
of the bending rigidity to a power-law
\begin{equation}
\kappa \sim  \frac{1}{|{\bf q}|^\eta}
\label{ab}
\end{equation}
The divergence of ${\rm Im} (\kappa({\bf q}, 0))$ at vanishing momentum can be taken as a 
consequence of the asymptotic behavior (\ref{ab}), as long as 
the ratio between the real and the imaginary part of the bending rigidity has a finite 
limit as ${\bf q} \rightarrow 0$.

\begin{figure}[h]
\begin{center}
\mbox{\epsfxsize 4.8cm \epsfbox{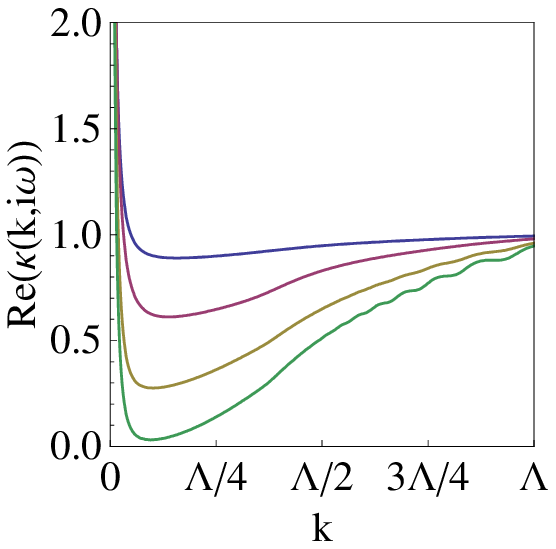} \hspace{3cm}
  \epsfxsize 4.8cm \epsfbox{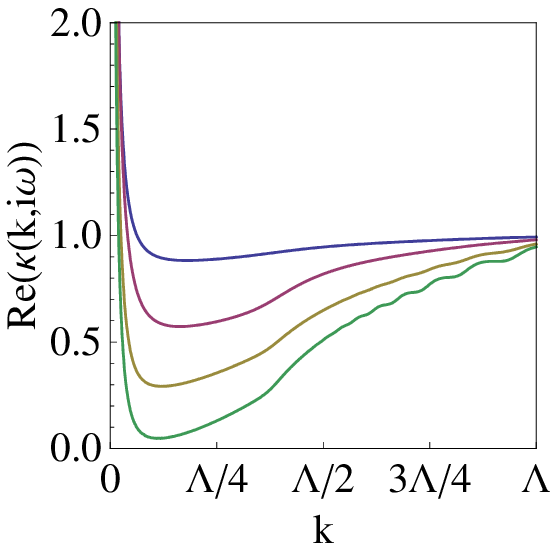}}  \\
 \hspace{0.3cm}  (a) \hspace{6.6cm} (b) 
\end{center}
\caption{Plot of the real part of the effective bending rigidity for (a) $2\pi T = 10$ meV, and (b) $2\pi T = 20$ meV. The curves correspond in both figures to a frequency $\omega = 0.25$ eV and values of the deformation potential $g = 28.0, 28.7, 28.84, 28.87$ eV (from top to bottom).}
\label{two}
\end{figure}

Finally, for a sufficiently large value of the deformation potential, there is a point 
where the real part of the bending rigidity $\kappa({\bf q}, i\overline{\omega}_q)$ vanishes, 
marking the threshold beyond which it becomes negative for an increasingly wide range of 
frequencies and momenta. The vanishing of ${\rm Re} (\kappa({\bf q}, i\overline{\omega}_q))$ 
starts taking place at a frequency $\overline{\omega}_q \sim 0.1$ eV, and it points at the transition 
to a different phase of the system as one cannot make physical sense of negative values of 
the bending rigidity. The approach to the critical point is shown in Fig. \ref{two} for two 
different values of the temperature. The critical value $g_c$ of the deformation potential 
depends slightly on $T$, and it is placed always close but below the upper limit marking the 
instability of the in-plane elastic deformations.

Negative values of the real part of $\kappa({\bf q}, i\overline{\omega}_q)$
point at the transition to a phase that cannot be cast in terms of the original phonon fields. The 
vanishing of the bending rigidity at the critical point implies the appearance of very soft
modes which, in turn, make the system more prone to develop some sort of spontaneous breakdown of 
symmetry with condensation of flexural phonons. The investigation of this effect requires however a
different approach to discern the possible divergence of any of the susceptibilities of the field 
theory, a question that we address specifically in the next section.

\section{Condensation of flexural phonons}

In the model considered above, the effective bending rigidity has a bounce as a function of the momentum for large electron-phonon coupling. This is a consequence of the nontrivial behavior of the coupling function $K({\bf q})$, which has different attractive and repulsive regimes. The attraction between flexural phonons is anyhow the actual reason for the drastic decrease observed in the bending rigidity. For the sake of clarifying the nature of the corresponding instability, it is then convenient to carry out the investigation of a simpler model where the interaction does not change its character, having an attractive potential 
\begin{equation}
K({\bf q}) = - G
\label{const}
\end{equation}
with constant $G$.
We note that one can make physical sense of this model as an approximate description of a graphene layer under suitably large doping. In that situation, the electron-hole susceptibility may have a leading term proportional to the chemical potential $\mu $ (measured from the Dirac point), leading to $\chi ({\bf q}, \omega_q ) \sim - \mu /v_F^2$. It is feasible then that, at sufficiently large doping levels (with $\mu \sim 1$ eV), the coupling $K({\bf q})$ obtained from (\ref{kq}) may be dominated by a negative term in the relevant range of strong electron-phonon coupling.

If one applies now the self-consistent screening approximation, it can be seen that the model with the coupling (\ref{const}) has a critical point which is again characterized by the vanishing of the effective bending rigidity, taking place now at momentum ${\bf q} = 0$. The self-consistent resolution (keeping the phonon interaction unrenormalized at this point) leads to a critical value of the coupling $G$ that vanishes for large values of the cutoff $\Lambda $ in the momentum integrals as
\begin{equation}
G_c (\Lambda ) \sim \frac{\sqrt{\rho } \kappa_0^{3/2}}{\log (\Lambda )}
\label{gcrit}
\end{equation}  
This behavior leads to a vanishing $G_c$ in the continuum limit, which is consistent with the renormalization of the model that we discuss in what follows. 

The simplicity of the interaction (\ref{const}) makes possible to use powerful scaling methods for the characterization of the critical point. As observed at the end of Sec. II, the field theory of flexural phonons has a scale invariance that allows to encode the most important quantum corrections into the renormalization of a finite number of parameters (the bending rigidity $\kappa $ and the coupling $K$ in our case). In the scaling approach, a key distinction is made between the original bare theory, written with the set of unrenormalized parameters and integrals running down to a microscopic cutoff, and the renormalized theory, in which one has to be able to make predictions independent of the value of the cutoff\cite{amit}. The passage is done through the redefinition of the parameters
\begin{eqnarray}
\kappa_0  & = &  Z_\kappa   \kappa_R           \\
    K  & = &   Z_K   K_R
\end{eqnarray}
If the theory is well-behaved (renormalizable), it must be possible to absorb all the dependences on the microscopic cutoff into the factors $Z_\kappa$ and $Z_K$, allowing to express the observables of the theory just in terms of cutoff-independent parameters $\kappa_R$ and $K_R$.

To see how this program works in our model to large orders in perturbation theory, it is convenient to implement a dimensional regularization of the integrals instead of using a more conventional high-energy cutoff in momentum space. We start then by writing the action with a slight deviation of the spatial dimension from $D = 2$ to $D = 2 - \epsilon$, in such a way that (\ref{act}) becomes
\begin{eqnarray}
S_{\rm eff}  & = &   \frac{1}{2} \int \frac{d^D q}{(2\pi)^D} \frac{d\omega}{2\pi} \; (\rho \: \omega^2  -
Z_\kappa \kappa_R {\bf q}^4  ) h({\bf q}, \omega) \: h(-{\bf q}, -\omega)
                                                 \nonumber                  \\
  &  &  -  \frac{1}{8}  Z_K K_R \:  \mu^\epsilon  \int \frac{d^D q}{(2\pi)^D} \frac{d\omega}{2\pi}  \;
 P_{ij}^{(T)} ({\bf q}) \: h_{ij} ({\bf q}, \omega) \: P_{kl}^{(T)} ({\bf q}) \: h_{kl} ({-\bf q}, -\omega)
\label{actdr} 
\end{eqnarray}
$\mu $ being an auxiliary momentum scale. In this approach, the $\log (\Lambda )$ factors arising in the theory regularized with a momentum cutoff $\Lambda $ are replaced by $1/\epsilon $ poles, which are supposed to be absorbed into an appropriate definition of 
the factors $Z_\kappa$ and $Z_K$.

The critical point of the theory has in particular a reflection in the renormalization of the bending rigidity $\kappa_0$. This can be analyzed from the phonon self-energy corrections $\Sigma ({\bf q}, \omega_q)$ or, equivalently, studying the renormalization of the composite operator
\begin{equation}
\Phi ({\bf r}, t)  =  \nabla^2 h ({\bf r}, t)  \: \nabla^2 h ({\bf r}, t)
\label{comp}
\end{equation}
The vertex corresponding to the operator (\ref{comp}) is given by 
\begin{equation}
 \Gamma ({\bf q},\omega_q;{\bf k},\omega_k)  = 
   \langle  \Phi ({\bf q},\omega_q) 
        h ({\bf k} - {\bf q},\omega_k - \omega_q) 
           h (-{\bf k},-\omega_k) \rangle_{\rm 1PI}
\label{vert}
\end{equation}
where 1PI denotes that we take the one-particle irreducible part of the correlator. It can be easily seen, from inspection of the respective diagrammatic contributions, that the vertex (\ref{vert}) is related to the phonon self-energy through the equation
\begin{equation}
\frac{\partial \Sigma ({\bf k}, \omega_k) }{\partial  \kappa_0} =  \Gamma ({\bf 0},0;{\bf k},\omega_k) 
\label{ward}
\end{equation}
In order to study the scaling of the bending rigidity, we can therefore focus on $\Gamma ({\bf 0},0;{\bf k},\omega_k)$, which is in general easier to analyze in comparison to $\Sigma ({\bf k}, \omega_k)$.

The bare vertex (\ref{vert}) is not a finite quantity in the physical limit $\epsilon \rightarrow 0$, as it can be seen from (\ref{ward}) that it inherits the cutoff dependence of $\kappa_0$. We can rewrite that equation in the form
\begin{equation}
\frac{\partial \Sigma ({\bf k}, \omega_k) }{\partial  \kappa_R} =  \frac{\partial \kappa_0}{\partial \kappa_R} \Gamma ({\bf 0},0;{\bf k},\omega_k) 
\label{rward}
\end{equation}
The left-hand-side of Eq. (\ref{rward}) must be a finite quantity in the limit $\epsilon \rightarrow 0$, since it can be written as a function only of the cutoff-independent bending rigidity $\kappa_R$. This means that it must be possible to absorb all the $1/\epsilon $ poles of $\Gamma ({\bf 0},0;{\bf k},\omega_k)$ into a multiplicative renormalization of the vertex
\begin{equation}
\Gamma_{\rm ren} = Z_\Gamma \Gamma
\label{mult}
\end{equation}
with
\begin{equation}
Z_\Gamma = \frac{\partial \kappa_0}{\partial \kappa_R}
\label{zgam}
\end{equation}
leading to a finite vertex $\Gamma_{\rm ren}$ in the physical limit $D = 2$.

\begin{figure}[h]
\begin{center}
\raisebox{0.4cm}{\epsfxsize 3.2cm \epsfbox{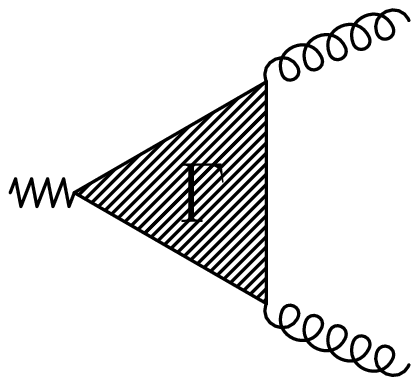}} \hspace{0.5cm} 
   \raisebox{1.7cm}{\Large $=$}
 \hspace{0.5cm} 
 \raisebox{0.85cm}{\epsfxsize 2.0cm \epsfbox{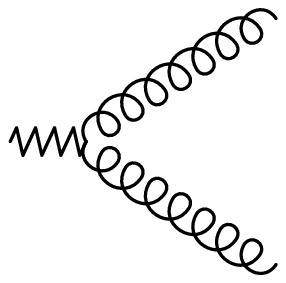}}
 \hspace{0.5cm}  \raisebox{1.7cm}{\Large $+$}  \hspace{0.5cm}
 \raisebox{0.05cm}{\epsfxsize 3.8cm \epsfbox{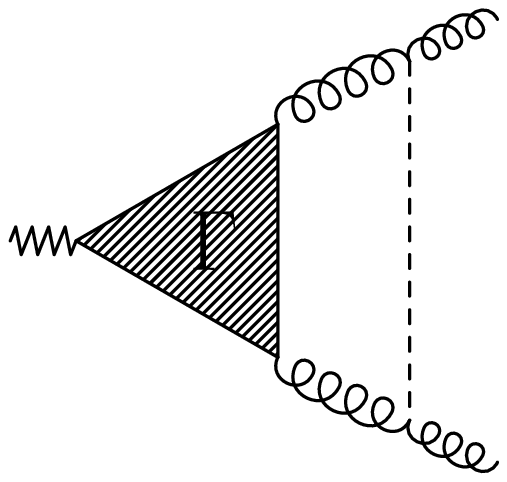}}  
\end{center}
\caption{Diagramatic equation defining the ladder approximation to the vertex with insertion of the operator $\nabla^2 h  \nabla^2 h$ (zig-zag line). The curly line represents the flexural phonon propagator and the dashed line corresponds to the flexural phonon interaction.}
\label{ladder}
\end{figure}

The identification of a possible critical point in the renormalized vertex requires the adoption of a nonperturbative approach to the many-body theory. For this purpose, we may resort to a most affordable ladder approximation for the computation of $\Gamma ({\bf 0},0;{\bf k},\omega_k)$, which is encoded in the self-consistent diagrammatic equation represented in Fig. \ref{ladder}. This can be expressed as 
\begin{equation}
i \Gamma ({\bf 0},0;{\bf k},\omega_k)  =  i {\bf k}^4  +  G \mu^\epsilon \int \frac{d^D p}{(2\pi)^2}  \frac{d\omega_p }{2\pi}
  \frac{({\bf k}^2 {\bf p}^2 - ({\bf k} \cdot {\bf p})^2)^2}{|{\bf k}-{\bf p}|^4} \: \Gamma ({\bf 0},0;{\bf p},\omega_p) \: \frac{1}{(\rho \omega_p^2 - \kappa_R \: {\bf p}^4 + i\eta)^2}
\label{lself}
\end{equation}
At this stage, the ladder approximation amounts to a sum of ladder diagrams that do not contain themselves any phonon self-energy loop. Therefore, it is safe to take the renormalized bending rigidity $\kappa_R$ in the phonon propagators in (\ref{lself}). The solution of the self-consistent equation can be expressed as a power series in the dimensionless coupling $\widetilde{G} \equiv G/\sqrt{\rho} \kappa_R^{3/2}$
\begin{equation}
\Gamma ({\bf 0},0;{\bf k},\omega_k) = 
 {\bf k}^4  +  {\bf k}^4 \sum_{n=1}^{\infty}    \widetilde{G}^n  \mu^{n\epsilon }  
                           \frac{s_n }{|{\bf k}|^{n\epsilon}} 
\label{ser}
\end{equation}
The coefficients $s_n$ diverge in the limit $\epsilon \rightarrow 0$ with powers of $1/\epsilon $, which can be absorbed into the multiplicative redefinition (\ref{mult}) with a renormalization factor that has the general structure
\begin{equation}
Z_\Gamma = 1 + \sum_{i=1}^{\infty} \frac{c_i (\widetilde{G} )}{\epsilon^i}
\label{pole}
\end{equation}

The factor $Z_\Gamma$ contains important information about the scaling of the renormalized vertex, which develops an anomalous dependence on the momentum scale $\mu $ leading to the long-wavelength behavior
\begin{equation}
\Gamma_{\rm ren} \sim  \left(  \frac{\mu }{|{\bf k}|}  \right)^{\gamma}
\label{scal}
\end{equation}
The original vertex $\Gamma $, as any quantity in the bare theory, does not know about the auxiliary scale $\mu $, so that the anomalous exponent $\gamma $ can be obtained as\cite{amit}
\begin{equation}
\gamma = \frac{\mu }{Z_\Gamma}  \frac{\partial Z_\Gamma}{\partial \mu}
\label{gam}
\end{equation}
On the other hand, the own independence of (\ref{ser}) on $\mu $ requires the implicit dependence of $\widetilde{G}$ on the auxiliary scale given by
\begin{equation}
\mu  \frac{d }{d \mu} (\widetilde{G} \mu^\epsilon )  =  0 
\end{equation}
Thus we get
\begin{equation}
\gamma = -\epsilon  \frac{\widetilde{G} }{Z_\Gamma} \frac{d Z_\Gamma}{d\widetilde{G} } 
\label{gam2}
\end{equation}
In principle, multiple poles from $Z_\Gamma $ may contribute to the right-hand-side of (\ref{gam2}), but they can cancel out provided that the hierarchy of equations
\begin{equation}
\frac{d}{d\widetilde{G} } c_{i+1} (\widetilde{G})
   - c_i (\widetilde{G}) \: \frac{d}{d\widetilde{G} } c_1 (\widetilde{G}) = 0
\label{cond}
\end{equation}
is satisfied. In that case, the anomalous dimension is given by\cite{ram} 
\begin{equation}
\gamma = -\widetilde{G} \frac{d}{d\widetilde{G} } c_1 (\widetilde{G})
\label{gamm}
\end{equation}
making it only dependent on the coupling $\widetilde{G} $.

It can be shown that the field theory of flexural phonons fulfills indeed the conditions (\ref{cond}) in the ladder approximation. 
Successive terms of the power series in (\ref{ser}) can be obtained recursively by using the formula 
\begin{equation}
G \mu^\epsilon \int \frac{d^D p}{(2\pi)^2}  \frac{d\overline{\omega}_p }{2\pi}
  \frac{({\bf k}^2 {\bf p}^2 - ({\bf k} \cdot {\bf p})^2)^2}{|{\bf k}-{\bf p}|^4} \: {\bf p}^4 \frac{\mu^{n\epsilon} }{|{\bf p}|^{n\epsilon}} \: \frac{1}{(\rho \overline{\omega}_p^2 + \kappa_R \: {\bf p}^4 )^2} = {\bf k}^4 \: \widetilde{G} \mu^{(n+1)\epsilon} \frac{A_n(\epsilon)}{|{\bf k}|^{(n+1)\epsilon}}
\end{equation}
with 
\begin{equation}
A_n(\epsilon) = \frac{3}{64\pi } (4\pi )^{\epsilon /2}   
  \frac{\Gamma \left(\tfrac{n+1}{2} \epsilon  \right) \Gamma \left(2 - \tfrac{n+1}{2}\epsilon \right) 
                                \Gamma \left(1 - \tfrac{1}{2}\epsilon \right) }
  {  \Gamma \left(1 + \tfrac{n}{2}\epsilon \right) \Gamma \left(3 - \tfrac{n+2}{2}\epsilon \right) }
\end{equation}
We get in this way the relation
\begin{equation}
s_{n+1} =  A_n(\epsilon) \: s_n
\label{recur}
\end{equation}
The different poles arising from the coefficients $s_n$ can be then canceled out in the renormalized vertex by adjusting appropriately the residues in (\ref{pole}). We can compute analytically for instance the first perturbative orders
\begin{eqnarray}
c_1 (\widetilde{G} )  & = & -\lambda \widetilde{G} -\frac{1}{8} \lambda^2 \widetilde{G}^2 -\frac{1}{24} \lambda^3 \widetilde{G}^3 -\frac{5}{256} \lambda^4 \widetilde{G}^4 -\frac{7}{640} \lambda^5 \widetilde{G}^5   
                                                   +    \ldots   \label{firstd}     \\
c_2 (\widetilde{G} ) & = &  \frac{1}{2} \lambda^2 \widetilde{G}^2 +\frac{1}{8} \lambda^3 \widetilde{G}^3 +\frac{19}{384} \lambda^4 \widetilde{G}^4 +\frac{19}{768}  \lambda^5 \widetilde{G}^5
                                                            +  \ldots                     \\
c_3 (\widetilde{G} ) & = &  -\frac{1}{6} \lambda^3 \widetilde{G}^3 -\frac{1}{16} \lambda^4 \widetilde{G}^4 -\frac{11}{384} \lambda^5 \widetilde{G}^5     + \ldots     \\
c_4 (\widetilde{G} ) & = &  \frac{1}{24} \lambda^4 \widetilde{G}^4 +\frac{1}{48} \lambda^5 \widetilde{G}^5     +  \ldots        \\
c_5 (\widetilde{G} ) & = &  -\frac{1}{120} \lambda^5 \widetilde{G}^5    +   \ldots
\label{lastd}
\end{eqnarray}
with $\lambda =  3/64\pi$. It can be seen that the analytic expressions in (\ref{firstd})-(\ref{lastd}) satisfy the conditions (\ref{cond}). A most important property is also that the residues $c_i$ do not contain any nonlocal dependence (in fact any dependence) on the momentum of the vertex, which guarantees that the renormalization only involves the redefinition of a finite number of local operators in the field theory.

\begin{figure}

\begin{center}
\mbox{\epsfxsize 7cm \epsfbox{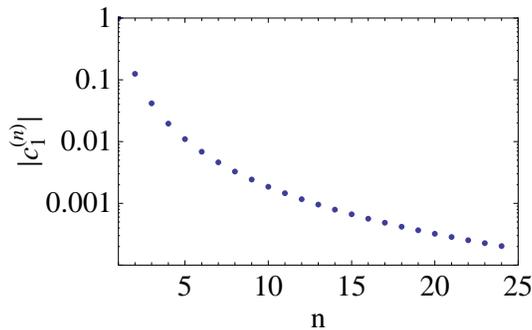}} 

\end{center}
\caption{Plot of the absolute value of the coefficients $c_1^{(n)}$ in the 
expansion of $c_1 (\widetilde{G} )$ as a power series of the effective 
coupling $\widetilde{G} $.}
\label{three}
\end{figure}

Furthermore, the computation of the residues can be carried out numerically to much larger orders in perturbation theory, which allows to inspect the behavior of $\gamma $ in the nonperturbative regime. Thus, we have been able to evaluate the residues $c_i$ to order $\widetilde{G}^{24}$, finding that the terms in each power series in the $\widetilde{G}$ coupling approach a geometric progression. We have paid particular attention to the residue $c_1$, which has an expansion 
\begin{equation}
c_1 (\widetilde{G}) = \sum_{n=1}^{\infty} c_1^{(n)}  \lambda^n \widetilde{G}^n
\label{c1}
\end{equation} 
with coefficients $c_1^{(n)}$ that approach a constant value at large $n$. The result of their numerical computation is represented in Fig. \ref{three}. A very precise fit of the dependence of the coefficients on the order $n$ can be indeed achieved by assuming the scaling 
\begin{equation}
\frac{c_1^{(n)}}{c_1^{(n-1)}} =  c + \frac{c'}{n}  +  \frac{c''}{n^2}  
    +   \frac{c'''}{n^3}  +   \ldots
\label{sca}
\end{equation} 
Using the results to order $\widetilde{G}^{24}$, we find that $c = 1.0, c' = -2.5, c'' = 1.5$, and values of $c''', c'''', \ldots$ that are at most of order $10^{-11}$. We may conclude therefore that the series (\ref{c1}) has a finite radius of convergence given by 
\begin{equation}
\widetilde{G}_c = \frac{1}{c\lambda}  = \frac{64\pi}{3}
\label{crit2}
\end{equation}

The singularity of the residue $c_1 (\widetilde{G})$ at $\widetilde{G}_c$ has important consequences from the physical point of view, since it implies the divergence of the anomalous dimension of the vertex according to (\ref{gamm}). In fact, any expectation value involving the operator (\ref{comp}) needs to be made cutoff-independent multiplying it by the factor $Z_\Gamma $, which means that the anomalous exponent $\gamma $ has to contribute to the scaling dimension of the correlator. The divergence of the susceptibility of the $\Phi $ field implies in particular that the composite operator must have a nonvanishing expectation value at $\widetilde{G}_c$, that is 
\begin{equation}
\langle   |\nabla^2 h ({\bf r}, t)|^2   \rangle   \neq  0
\end{equation}
It becomes clear that $\widetilde{G}_c$ represents a critical point of the many-body theory, with the pattern of symmetry breaking dictated by the condensation of the field $\Phi $.
  
It is now easy to see that the critical coupling $\widetilde{G}_c$ must correspond to the singular behavior characterized by the vanishing of the effective bending rigidity in the self-consistent screening approximation. From Eq. (\ref{zgam}) we have
\begin{equation}
Z_\kappa +  \frac{\partial Z_\kappa}{\partial \kappa_R} \kappa_R  =  Z_\Gamma
\end{equation}
We infer that $Z_\kappa $ must also diverge at the critical coupling $\widetilde{G}_c$. If we think of the relation between $\kappa_0$ and $\kappa_R$ as giving the correspondence between the values of the bending rigidity at short and long wavelengths, we conclude that any finite value of $\kappa_0$ at the microscopic scale must lead to a vanishing effective bending rigidity $\kappa_R$ when the coupling is set at $\widetilde{G}_c$. This argument highlights again that the origin of the critical behavior is the divergence of $\partial \kappa_0 / \partial \kappa_R$ at the critical interaction strength.

Regarding the precise location of the critical point, it has to be noted that the value given in (\ref{crit2}) does not admit a direct comparison with the result that is obtained in the self-consistent screening approximation. This includes the iterated effect of phonon-loop corrections in the own phonon self-energy, while these corrections are absent in our renormalization of the vertex. This can be certainly refined to incorporate the effect of phonon self-energy insertions in the ladder approximation. It can be shown that, including for instance one-loop phonon corrections in our computation of the vertex, the value of $c$ in (\ref{sca}) increases from 1 to 1.8. In general, iterating more and more phonon-loop corrections to the vertex leads to a significant reduction of the critical coupling with respect to the value quoted in (\ref{crit2}). The overall picture is then that the critical point is anyhow preserved after taking into account the phonon self-energy corrections, and that the behavior (\ref{gcrit}) may be consistent with a nonvanishing critical value $\widetilde{G}_c$ for the dimensionless renormalized coupling, as this is defined in terms of the effective bending rigidity $\kappa_R $ that must vanish at the critical point.

\section{Discussion}

We have studied the many-body theory of flexural phonons in a metallic membrane like graphene, where the interaction between the electronic excitations and elastic deformations may induce important effects. We have shown that, for sufficiently strong electron-phonon interaction, the theory has a critical point characterized by the vanishing of the effective bending rigidity $\kappa_R $ at long wavelengths. In this investigation, we have relied on a self-consistent screening approximation to evaluate the flexural phonon self-energy, while refining the analysis to incorporate the main features of the momentum dependence of the electron-phonon couplings.

We have seen that the effective bending rigidity has in general a bounce as a function of momentum, which is a consequence of the interplay between the electron-induced attraction and the natural repulsive interaction between flexural phonons. This investigation has clarified that the instability in the sector of flexural phonons takes place without the development of an in-plane static distortion of the lattice, which is avoided due to the significant reduction of the electron-phonon couplings for in-plane phonons at large momenta.  

We have applied a renormalization group analysis to identify the order parameter characterizing the new phase marked by the vanishing of the effective bending rigidity. In a model with constant attractive interaction between flexural phonons, we have shown that there is a close relation between the renormalization factors of the bending rigidity and the mean curvature field $\nabla^2 h$. This implies that the vanishing of $\kappa_R$ and the onset of a nonzero expectation value for the mean curvature can be viewed as different manifestations of the same effect. We may therefore conclude that the critical point we have studied corresponds to a rippling transition at which the flat geometry becomes an unstable configuration of the metallic membrane.   

As $\kappa_R \rightarrow 0$ at the critical point, the weight of the kinetic energy of the membrane vanishes, which makes it more susceptible to an external influence and, in particular, to tension exerted by external forces. This is the route to symmetry breaking that was actually investigated in Ref. \cite{prl}, finding that any slight amount of negative tension may turn the flat geometry unstable, inducing a non-vanishing expectation value of the field $\partial_i h $. In that analysis, there was however no conclusive evidence of the spontaneous out-of-plane distortion of the membrane in the limit situation with vanishing tension. The results presented here imply that, in the absence of tension, the instability at vanishing $\kappa_R $ induces anyhow the condensation of the mean curvature field $\nabla^2 h$. The same conclusion applies to the case of very small positive or negative tension, when leading to modulations of $\partial_i h $ with a period larger than the critical length scale marking the vanishing of the bending rigidity, since the presence of such a small perturbation does not invalidate in that case the results of Sec. III.   

It has to be also pointed out that the vanishing of $\kappa_R $ leads to a regime of strong coupling which could trigger the divergence of a different susceptibility apart from that of the mean curvature $\nabla^2 h$. In this respect, an alternative possibility corresponds to the other scale-invariant operator in the action (\ref{actdr}), built from the field $P_{ij}^{(T)} \partial_i h \partial_j h$. The condensation of this field falls again in the route to symmetry breaking pursued in Ref. \cite{prl}, and it has been also recently investigated in Ref. \cite{gdw}. As $\kappa_R \rightarrow 0$, the propagator of $P_{ij}^{(T)} \partial_i h \partial_j h$ can develop a pole in a RPA framework, which has to be interpreted as the signal of an out-of-plane static distortion given by the modulation of that field. 

The strong-coupling character of the regime with $\kappa_R \rightarrow 0$ does not allow our renormalization group method to discern the prevalence of any of the two symmetry-breaking patterns driven respectively by $\nabla^2 h$ and $P_{ij}^{(T)} \partial_i h \partial_j h$. We note anyhow that the effects deriving from the condensation of each of these fields have to be very different in the two cases. The condensation of $P_{ij}^{(T)} \partial_i h \partial_j h$ actually implies a non-vanishing expectation value of $\partial_i h$ which, being a vector, may induce vorticity in the geometry of the membrane. On the other hand, the condensation of the field $\nabla^2 h$ will lead in general to a smooth configuration of the out-of-plane distortion. In a real material both patterns of symmetry breaking may be present. A careful analysis of experimental data of the curvature distributions may be however required to disentangle the two contributions, for the purpose of assessing the relative significance of each of them in the geometry of the real metallic membrane.

\section*{Acknowledgments}
I thank F. Guinea and P. San-Jos\'e for very useful discussions.
The financial support from MICINN (Spain) through grant FIS2011-23713 is 
gratefully acknowledged.

\appendix*

\section{Tight-binding approach to electron-phonon couplings}

In the tight-binding approach, the expression of the electron-phonon couplings can be obtained from the variation of the potential energy of the electron system under the displacement of the atoms in the underlying lattice. We take as starting point the expectation value of the periodic potential formed by superposition of atomic potentials $v({\bf r})$ centered at the different lattice sites ${\bf R}_{n\alpha}$, labelled by unit cell $n = 1, \ldots N$ and the position $\alpha $ within the unit cell,
\begin{equation}
V ({\bf r}) = \sum_{n\alpha } v({\bf r} - {\bf R}_{n\alpha })
\end{equation}
We can then evaluate the potential energy for electronic states with well-defined momentum ${\bf k}$, represented by a Bloch superposition of wavefunctions $\phi ({\bf r})$ for the atomic orbitals
\begin{equation}
\Psi_{\bf k} ({\bf r}) = \frac{1}{\sqrt{N}} \sum_{\alpha } \psi_{\alpha }({\bf k})
   \sum_n e^{i {\bf k} \cdot {\bf R}_{n\alpha } } 
                 \phi ({\bf r} - {\bf R}_{n\alpha })
\end{equation}
We end up in this way with the matrix element
\begin{equation}
E_{{\bf k}',{\bf k}} \equiv \langle \Psi_{{\bf k}'}  | V |  \Psi_{\bf k}  \rangle 
 =  \frac{1}{N} \sum_{\alpha,\alpha'} \psi^*_{\alpha' }({\bf k}') 
   \psi_{\alpha }({\bf k}) \sum_{n,n'}  
  e^{i (- {\bf k}' \cdot {\bf R}_{n'\alpha' } +{\bf k} \cdot {\bf R}_{n\alpha }) }
 v(n'\alpha', n\alpha)
\label{var}
\end{equation}
written in terms of the atomic matrix element
\begin{equation}
v(n'\alpha', n\alpha) = 
 \int d^2 r \; \phi ({\bf r} - {\bf R}_{n'\alpha' }) 
  \sum_{n''\alpha''}  v({\bf r} - {\bf R}_{n''\alpha'' })
   \phi ({\bf r} - {\bf R}_{n\alpha }) 
\label{at}
\end{equation}

Under a vibration of the lattice, the displacement of the atoms from their equilibrium positions gives rise to a so-called deformation potential. An approximate expression for the variation of $E_{{\bf k}',{\bf k}}$ can be obtained by truncating the sums in (\ref{var}) and (\ref{at}) to the terms in which the points ${\bf R}_{n\alpha }, {\bf R}_{n'\alpha' }$ and ${\bf R}_{n''\alpha'' }$ have some degree of proximity. In this respect, one usually makes a distinction between the on-site and the off-site deformation potential, corresponding respectively to taking ${\bf R}_{n\alpha } = {\bf R}_{n'\alpha' }$ on one hand, and either ${\bf R}_{n\alpha } = {\bf R}_{n''\alpha'' }$ or ${\bf R}_{n'\alpha' } = {\bf R}_{n''\alpha'' }$ on the other. In general, the component of the electron-phonon coupling arising from the on-site deformation potential is much larger than that from the off-site counterpart. For this reason, we concentrate in what follows on the on-site contribution.  

Focusing then on the case with ${\bf R}_{n\alpha } = {\bf R}_{n'\alpha' }$, we can write the matrix element (\ref{at}) as
\begin{equation}
v(n\alpha, n\alpha) = 
 \int d^2 r \; \phi ({\bf r} ) 
  \sum_{n''\alpha''}  v({\bf r} - {\bf R}_{n''\alpha'' } + {\bf R}_{n\alpha })
   \phi ({\bf r} ) 
\label{at2}
\end{equation}
The deviation of the atoms from the equilibrium position, ${\bf R}_{n\alpha } \rightarrow {\bf R}_{n\alpha } + {\bf d} ({\bf R}_{n\alpha })$,
leads to the on-site deformation potential
\begin{eqnarray}
  \Delta v_{\rm on}    &  =  &
 - \int d^2 r \; \phi ({\bf r} )    \sum_{n''\alpha''}
 \mbox{\boldmath $\nabla $} v({\bf r} - {\bf R}_{n''\alpha'' } + {\bf R}_{n\alpha }) 
 \cdot ( {\bf d} ({\bf R}_{n''\alpha'' }) - {\bf d} ({\bf R}_{n\alpha })  )
  \; \phi ({\bf r}                     )                  \nonumber      \\
   &  &    +  \frac{1}{2} \int d^2 r \; \phi ({\bf r})    \sum_{n''\alpha''}
  (  {\bf d} ({\bf R}_{n''\alpha'' }) - {\bf d} ({\bf R}_{n\alpha })  )\cdot                
   \mbox{\boldmath $\nabla $}  
  \mbox{\boldmath $\nabla $} v({\bf r} - {\bf R}_{n''\alpha'' } + {\bf R}_{n\alpha }) 
 \cdot ( {\bf d} ({\bf R}_{n''\alpha'' }) - {\bf d} ({\bf R}_{n\alpha })  )
  \;  \phi ({\bf r})     +   \ldots                 \;\;\;\;
\label{von}
\end{eqnarray}

We are interested in particular in the situation where the displacement ${\bf d} ({\bf R}_{n\alpha })$ arises from a phonon of the lattice. Then, such a field can be represented in terms of the polarization ${\bf e}_{\alpha } ({\bf q})$ of the given phonon mode as
\begin{equation}
{\bf d} ({\bf R}_{n\alpha }) = 
   e^{i {\bf q} \cdot {\bf R}_{n\alpha } }  {\bf e}_{\alpha } ({\bf q})
\end{equation}
Inserting this expression into (\ref{von}) and the atomic matrix element back into (\ref{var}), we arrive at the electron-phonon couplings for the respective interactions with one and two phonon fields:
\begin{eqnarray}
 g_{\alpha, \alpha }^{(1)} ({\bf k}; {\bf q})  
   &  =  &    \psi^*_{\alpha }({\bf k} + {\bf q }) \psi_{\alpha }({\bf k})  
   \sum_{n'\alpha'}  
( e^{i {\bf q}\cdot {\bf u}_{n'\alpha' } } {\bf e}_{\alpha' } ({\bf q }) 
   - {\bf e}_{\alpha } ({\bf q })  )  \cdot
    {\bf n} ({\bf u}_{n'\alpha' })                     \label{oneph}    \\
 g_{\alpha, \alpha }^{(2)} ({\bf k}; {\bf q}, {\bf q' })  
  &  =  &   \psi^*_{\alpha }({\bf k} + {\bf q } + {\bf q' }) \psi_{\alpha }({\bf k})  
                                                        \nonumber        \\  
    &     &     \sum_{n'\alpha'} 
( e^{i {\bf q} \cdot {\bf u}_{n'\alpha' } } {\bf e}_{\alpha' } ({\bf q } ) 
    - {\bf e}_{\alpha } ({\bf q } )  ) \:
   {\bf m} ({\bf u}_{n'\alpha' }) \: 
( e^{i {\bf q}' \cdot {\bf u}_{n'\alpha' } } {\bf e}_{\alpha' } ({\bf q }') 
   -  {\bf e}_{\alpha } ({\bf q }')  )
\label{twoph}
\end{eqnarray}
where ${\bf u}_{n'\alpha' } = {\bf R}_{n'\alpha' } - {\bf R}_{n\alpha }$ and 
\begin{eqnarray}
n_a ({\bf u})  & = &
  \int d^2 r \;  \phi ({\bf r})  
    \nabla_a v({\bf r} - {\bf u})  \phi ({\bf r})  \nonumber  \\ 
m_{a b } ({\bf u})  & = &
   \int d^2 r \;  \phi ({\bf r})  
   \nabla_a  \nabla_b  v({\bf r} - {\bf u})  \phi ({\bf r}) 
\end{eqnarray}

We can now specialize the expressions (\ref{oneph}) and (\ref{twoph}) to the instances which are relevant for our study. First, there is the case of the in-plane acoustic longitudinal phonons, for which the polarization is simply given by
\begin{equation}
{\bf e}_{\alpha } ({\bf q}) = \frac{{\bf q}}{|{\bf q}|}
\end{equation}
One can easily see that the vector ${\bf n} ({\bf u}_{n'\alpha' })$ has to point in the same direction of the vector ${\bf u}_{n'\alpha' }$ \cite{pietro}. Then, we have for this type of phonons 
\begin{equation}
 g_{\alpha, \alpha }^{(1)} ({\bf k}; {\bf q})  
     \sim      \psi^*_{\alpha }({\bf k} + {\bf q }) \psi_{\alpha }({\bf k})  
   \sum_{n'\alpha'}  
( e^{i {\bf q}\cdot {\bf u}_{n'\alpha' } } - 1) \: {\bf q} \cdot {\bf u}_{n'\alpha' }
          \:  s (|{\bf u}_{n'\alpha' }|)
\label{in}
\end{equation}

\begin{figure}[h]
\begin{center}
\mbox{\epsfxsize 4.8cm \epsfbox{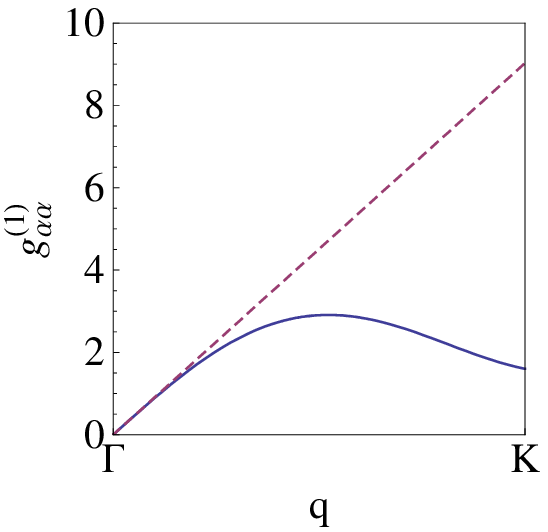} \hspace{3cm}
  \epsfxsize 4.8cm \epsfbox{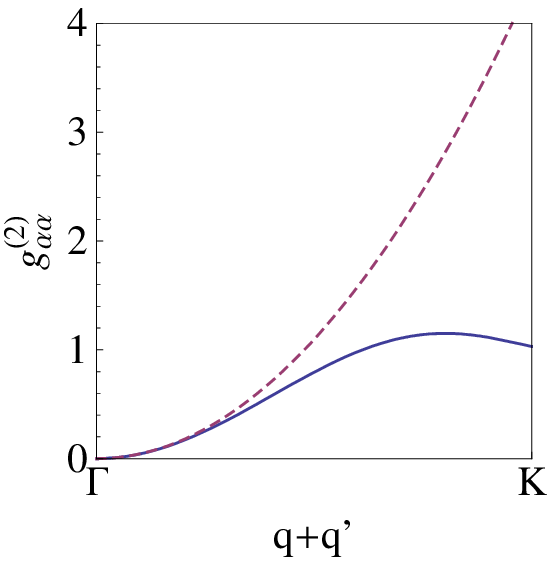}}  \\
 \hspace{0.3cm}  (a) \hspace{6.6cm} (b) 
\end{center}
\caption{Plot of the electron-phonon couplings obtained from the on-site deformation potential in the case of (a) in-plane acoustic longitudinal phonons, and (b) out-of-plane acoustic phonons (for phonon momenta ${\bf q } = {\bf q' }$). The dashed lines represent the respective linear and quadratic behaviors extrapolated from the long-wavelength limit.}
\label{appf1}
\end{figure}

In practice, one has to restrict the sum in (\ref{in}) to a certain set of neighbors of the point ${\bf R}_{n\alpha }$ when computing the electron-phonon coupling. At small ${\bf q }$, $g_{\alpha, \alpha }^{(1)}$ has a leading linear dependence on the momentum, which is in agreement with the form of the coupling in the continuum limit shown in (\ref{acte}). As the momentum increases, however, the electron-phonon coupling deviates significantly from the linear behavior, leading to a drastic reduction from what one could expect according to the linear small-${\bf q }$ dependence. This is illustrated in Fig. \ref{appf1}(a), where we have plotted the coupling (\ref{in}) including up to third nearest-neighbor contributions. The decreasing trend of $g_{\alpha, \alpha }^{(1)}$ at large ${\bf q }$ is consistent with the very small values predicted by detailed calculations of the coupling for in-plane acoustic phonons at the $K$ point of the Brillouin zone\cite{pisc}.

We shift now to the case of the out-of-plane acoustic phonons. These have a polarization 
\begin{equation}
{\bf e}_{\alpha } ({\bf q}) = {\bf z}
\end{equation}
where ${\bf z}$ stands for the unit vector in the direction transverse to the reference plane of the lattice. Then, it is clear that $g_{\alpha, \alpha }^{(1)}$ vanishes in the case of a free-standing membrane, as ${\bf n} ({\bf u}_{n'\alpha' })$  is orthogonal to ${\bf z}$ , and we have to consider the two-phonon coupling $g_{\alpha, \alpha }^{(2)}$. The tensor ${\bf m} ({\bf u}_{n'\alpha' })$ has certainly a nonvanishing $zz$ component, so that the evaluation of (\ref{twoph}) leads in this case to
\begin{equation}
 g_{\alpha, \alpha }^{(2)} ({\bf k}; {\bf q}, {\bf q' })  
  \sim  \psi^*_{\alpha }({\bf k} + {\bf q } + {\bf q' }) \psi_{\alpha }({\bf k})  
     \sum_{n'\alpha'}  ( e^{i {\bf q} \cdot {\bf u}_{n'\alpha' } } - 1) \:
                 ( e^{i {\bf q}' \cdot {\bf u}_{n'\alpha' } } - 1)
           \:   t (|{\bf u}_{n'\alpha' }|)
\label{oop}
\end{equation}

One can check that (\ref{oop}) has a leading quadratic dependence at low phonon momenta, proportional to ${\bf q} \cdot {\bf q}'$. This is again in agreement with the form of the coupling to out-of-plane displacements in the continuum limit, 
as expressed in the action (\ref{acte}). At large values of the momenta, there is however a marked deviation from the quadratic behavior, that leads the coupling $g_{\alpha, \alpha }^{(2)}$ to reverse the increasing trend when approaching the boundary of the Brillouin zone. This is appreciated in the plot of Fig. \ref{appf1}(b), which shows the coupling for momenta ${\bf q } = {\bf q' }$ along the $\Gamma K$ line.

The main conclusion of this analysis is that the different electron-phonon couplings building the dependence of $K({\bf q})$ on the deformation potential do not have the same values away from the continuum limit. The first $g^2$ dependence at the right-hand-side of (\ref{kq}) comes for instance from the coupling of flexural phonons through the exchange of the electron-hole polarization. The last fraction in that equation accounts instead for the exchange of in-plane longitudinal phonons, which are also coupled in turn to the electron-hole pairs. For momenta that are not in general small, it is then more accurate to express the coupling function as 
\begin{equation}
K({\bf q}) = 2\mu + \lambda - g_{\rm out}^2 |{\bf q}|/4v_F
 - \frac{(\lambda - g_{\rm in} g_{\rm out}  |{\bf q}|/4v_F)^2 }
                               {2\mu + \lambda - g_{\rm in}^2 |{\bf q}|/4v_F }
\label{kqgg}
\end{equation}
with deformation potentials $g_{\rm in}({\bf q}), g_{\rm out}({\bf q})$ that are equal at ${\bf q} \rightarrow 0$, but have to be in general extracted according to the behavior of the in-plane and out-of-plane electron-phonon couplings shown above. This is the approach that we have adopted in our study of the many-body theory of flexural phonons, applying it in particular to the resolution of the self-consistent screening approximation. 

The behavior at large momenta of the coupling to in-plane acoustic phonons gives in particular the clue of why the metallic membrane does not show an in-plane instability for relatively large values of the nominal deformation potential (as measured in the continuum limit). A momentum-independent deformation potential would make possible to reach the pole in the fraction of $K({\bf q})$ for quite low values of $g$, implying a spontaneous static distortion in the sector of in-plane phonons. Taking into account the reduction of the coupling $g_{\alpha, \alpha }^{(1)}$ shown above at large momenta, one can estimate instead that an in-plane instability of the membrane may not likely appear up to nominal values of the deformation potential (that is the limit $g_{\rm in}({\bf q} \rightarrow 0)$) as large as $g \approx 29$ eV. Before that point is reached, the attractive component of $K({\bf q})$ may become however quite large, explaining that the instability for out-of-plane deformations of the membrane may happen first, as shown in the paper.


\begin{thebibliography}{99}
\bibitem{fn}
This behavior can be obtained for instance from the resolution of the scaling
equations derived in Ref. \cite{prl}.
\end{thebibliography}
\end{document}